\documentclass[letterpaper, 10 pt, conference]{ieeeconf}  

\IEEEoverridecommandlockouts                              

\usepackage{amsmath} 
\usepackage{bm} 
\usepackage{graphicx}
\usepackage{caption}
\usepackage{makecell}
\usepackage{tabularx}
\usepackage{diagbox}
\usepackage{cite}
\usepackage{graphicx,subcaption}
\usepackage{booktabs}
\usepackage{breqn}
\usepackage{xcolor}

\newcommand{\highlight}[1]{%
  \colorbox{red!50}{$\displaystyle#1$}}

\begin{document}

\title{\LARGE \bf Sparse Identification of Lagrangian for Nonlinear Dynamical Systems via Proximal Gradient Method}

\author{Adam Purnomo and Mitsuhiro Hayashibe
\thanks{The authors are with the Neuro-Robotics Lab, Department of Robotics, Tohoku University, 980-8579, Sendai, Japan (email : {\tt\small adam.syammas.zaki.purnomo.p8@@dc.tohoku.ac.jp; hayashibe@tohoku.ac.jp})}}%

\maketitle

\begin{abstract}
Distilling physical laws autonomously from data has been of great interest in many scientific areas. The sparse identification of nonlinear dynamics (SINDy) and its variations have been developed to extract the underlying governing equations from observation data. However, SINDy faces certain difficulties when the dynamics contain rational functions. The principle of the least action governs many mechanical systems, mathematically expressed in the Lagrangian formula.  Compared to the actual equation of motions, the Lagrangian is much more concise, especially for complex systems, and does not usually contain rational functions for mechanical systems. Only a few methods have been proposed to extract the Lagrangian from measurement data so far. One of such methods, Lagrangian-SINDy, can extract the true form of Lagrangian of dynamical systems from data but suffers when noises are present.  In this work, we develop an extended version of Lagrangian-SINDy (xL-SINDy) to obtain the Lagrangian of dynamical systems from noisy measurement data. We incorporate the concept of SINDy and utilize the proximal gradient method to obtain sparse expressions of the Lagrangian. We demonstrated the effectiveness of xL-SINDy against different noise levels with four nonlinear dynamics: a single pendulum, a cart-pendulum, a double pendulum, and a spherical pendulum.  Furthermore, we also verified the performance of xL-SINDy against SINDy-PI (parallel, implicit), a recent robust variant of SINDy that can handle implicit dynamics and rational nonlinearities. Our experiment results show that xL-SINDy is 8-20 times more robust than SINDy-PI in the presence of noise. 
\end{abstract}

%
\IEEEpeerreviewmaketitle

\section{Introduction and Related Works}\label{sec1}
{S}{ince} the early modern history of humanity, scientists have always been trying to come up with models that can capture real-world phenomena. Such models are desired because they can be used to devise solutions to real-world problems. For centuries, the process of refining hypothesis and models from observation data have been conducted manually. Automating this process has long been of great interest in the scientific community. 


Many attempts have been made to extract physical laws autonomously from data. With the abundance of data and cheaper yet powerful hardware, the deep learning-based methods have gained a lot of attraction and have been widely used to model and control dynamical systems \cite{Battaglia2016,lenz2015}. It has also been shown that deep learning is also capable of approximating invariant quantities from dynamical systems such as the Hamiltonian \cite{HNN} and the Lagrangian \cite{LNN}. However, deep learning models act as black boxes; it does not provide insights into how each observation variable affects and relates to each other.


Recent trends, however, favor parsimonious models, models with the lowest complexity to describe the observation data. A ground-breaking work done by Schmidt and Lipson \cite{lipson2009} shows us that it is possible to extract the governing mathematical expressions from observation data. Symbolic regression is used to find the nonlinear differential equations that describe the behavior of the system, but symbolic regression tends to be expensive.  The sparse identification of nonlinear dynamics (SINDy) \cite{Brunton2016} models nonlinear differential equations of dynamics as a linear combination of nonlinear candidate functions and obtains a parsimonious model through sparse regression. 


While there are many applications of SINDy across different fields\cite{Sorokina2016, Dam2017, Loiseau2018, Brian2020}, SINDy faces certain difficulties when the dynamics contain rational functions. Including rational functions into the library of candidate functions would tremendously increase the size of the library, making the sparse regression challenging. A modification of SINDy, implicit-SINDy \cite{Mangan2016}, reformulates the SINDy problem into implicit form to address this challenge, albeit this method is sensitive to noise. SINDy-PI \cite{Kaheman2020} is proposed to improve the performance of implicit-SINDy in terms of noise robustness. While SINDy-PI is much more robust than implicit-SINDy, it can only obtain the correct dynamical structure with noise magnitude on the scale of up to $10^{-3}$  which might not be sufficient for a real-world application. On top of that, if the incorrect combination of denominator terms is discovered, the predicted system might blow up when the denominator is equal to zero

The principle of least action is fundamental to many dynamical systems \cite{leonhard1744}. The principle states the trajectory chosen by the system is the one that minimizes a certain cost function. This cost function is the so-called 'action,' which is defined as the integral of the Lagrangian for an input evolution between a certain time period. Compared to the underlying differential equations, Lagrangian has a desirable property in which it is a single scalar quantity that contains all information to predict the behavior of the systems. In robotics, the derivation of the dynamics often starts from the Lagrangian of the systems. 

Several works have been proposed to approximate the Lagrangian from data with polynomial basis functions \cite{hills2015, Ahmadi2018}. However, approximating Lagrangian with polynomial basis functions will only be useful for a particular trajectory of the system and will not likely generalize well across different initial conditions. Lagrangian-SINDy \cite{Hoang2020} is a SINDy-based method designed to extract the Lagrangian of nonlinear dynamics and is shown to be able to retrieve the true form of Lagrangian of several dynamical systems. However, the author mentioned in the paper that Lagrangian-SINDy is sensitive to noise and cannot recover the Lagrangian when the training data is corrupted by Gaussian noise even with magnitude in the scale of $10^{-7}$. Noise will always present in real-world systems, and developing a robust method against noise is important for real-world applications.

In this work, we propose a method called extended Lagrangian-SINDy (xL-SINDy) that can discover the true form of the Lagrangian and is more robust in the presence of noise compared to Lagrangian-SINDy and SINDy-PI. We demonstrated the effectiveness of xL-SINDy against different noise levels and compare the robustness of xL-SINDy against SINDy-PI in physical simulations with four dynamical systems: A single pendulum, a cart-pendulum, a double pendulum, and a spherical pendulum. This paper is organized as follows; section II describes how we use basic ideas from previous works to formulate the problem and develop the learning method, section III presents the results of simulation experiments on the aforementioned dynamical systems, and section IV provides closing and remarks.

\section{Method}\label{sec2}
\subsection{Problem Formulation}\label{subsec21}
Inspired by the concept of SINDy \cite{Brunton2016}, we consider a Lagrangian expression in a structure of a linear combination of nonlinear candidate functions. Let $\bm{q} = (q_1, q_2, ..., q_n)$ be the configuration of a system in a generalized coordinate of a system, the Lagrangian of the system is expressed as 

\begin{align}
&\mathcal{L} = \sum_{k=1}^{p} c_k\phi_{k}(\bm{q},\bm{\dot{q}}),
\end{align}
where, $\phi_{k}(\bm{q},\bm{\dot{q}})$, $k=1,...,p$ are a set of nonlinear candidate functions, and $c_{k}$, $k=1,...,p$ are the corresponding coefficients. We are interested to find the value of $\bm{c} = (c_1, c_2, ..., c_p)$ where we believe that the majority of the coefficients are zero. The Lagrangian of the system satisfies the Euler-Lagrange equations given by

\begin{align}
&\bm{\tau}_{ext} =  \frac{d}{dt} \nabla_{\bm{\dot{q}}}\mathcal{L} - \nabla_{\bm{q}}\mathcal{L},
\end{align}
where $\left(\nabla_{\bm{q}}\right)_i \equiv \frac{\partial}{\partial q_i}$. We consider three different scenarios: 

\begin{itemize}
    \item Case I: External input $\bm{\tau}_{ext}$ of the system is provided. 
    \item Case II: No external input is provided.
    \item Case III: Prior Lagrangian knowledge of a simpler system that forms a constituent of the system is provided.
\end{itemize}

\subsubsection{With External Input}
In the case where input $\bm{\tau}_{ext}$ is provided, substituting (1) in (2) yields
\begin{align}
&\bm{\tau}_{pred} = \frac{d}{dt}\sum_{k=1}^{p}c_k \nabla_{\bm{\dot{q}}}\phi_{k} - \sum_{k=1}^{p}c_k \nabla_{\bm{q}}\phi_{k},
\end{align}
where $\bm{\tau}_{pred}$ is the predicted value of the external input $\bm{\tau}_{ext}$ given a set of coefficient $\bm{c} = (c_1, c_2, ..., c_p)$. We can further expand the time derivative $\frac{d}{dt}$ by using chain rule, giving us the terms $\bm{\dot{q}}$ and $\bm{\ddot{q}}$

\begin{align}
\begin{split}
\bm{\tau}_{pred}&=\left( \sum_{k=1}^{p}c_k  \nabla^\top_{\bm{\dot{q}}}\nabla_{\bm{\dot{q}}}\phi_{k}\right) \bm{\ddot{q}} + \left( \sum_{k=1}^{p}c_k  \nabla^\top_{\bm{q}}\nabla_{\bm{\dot{q}}}\phi_{k}\right) \bm{\dot{q}} \\ &-\left(\sum_{k=1}^{p}c_k \nabla_{\bm{q}}\phi_{k}\right) \\ 
&= \sum_{k=1}^{p}c_k\left( \nabla^\top_{\bm{\dot{q}}}\nabla_{\bm{\dot{q}}}\phi_{k}\bm{\ddot{q}} + \nabla^\top_{\bm{q}}\nabla_{\bm{\dot{q}}}\phi_{k}\bm{\dot{q}} - \nabla_{\bm{q}}\phi_{k} \right) .
\end{split}
\end{align}
To avoid verbose notation, we define the following notations
\begin{align}
&\mathbf{M}_{k} = \nabla^\top_{\bm{\dot{q}}}\nabla_{\bm{\dot{q}}}\phi_{k},\\
&\mathbf{N}_{k} = \nabla^\top_{\bm{q}}\nabla_{\bm{\dot{q}}}\phi_{k},\\
&\mathbf{O}_{k} = \nabla_{\bm{q}}\phi_{k}.
\end{align}
Substituting (5), (6), and (7) in (4) yields
\begin{align}
\bm{\tau}_{pred} = \sum_{k=1}^{p}c_k\left(\mathbf{M}_{k}\bm{\ddot{q}} + \mathbf{N}_{k}\bm{\dot{q}} - \mathbf{O}_{k} \right),
\end{align}
and we define the following cost function that we want to minimize to obtain the Lagrangian of the system

\begin{align}
J(\bm{c}) = \| \bm{\tau}_{ext} - \bm{\tau}_{pred}(\bm{c}) \|_{2}^{2}.
\end{align}

We can use the above cost functions if the external torque is provided. In the case of passive systems where no external torque is provided, we will just end up minimizing the residual cost function $J(\bm{c}) = \|  - \bm{\tau}_{pred}(\bm{c}) \|_{2}^{2}$. From equation (8), we can observe that the $\bm{\tau}_{pred}(\bm{c})$ is in the form of linear combination of coefficient $\bm{c}$. Thus, minimizing this residual cost function is equivalent to the problem of finding a sparse null space which is an arduous task with current optimization methods. This will lead us to the formulation of a new cost function that will be explained in the next section. 

\subsubsection{Without External Input}
In the case of passive systems, in which no external input $\bm{\tau}_{ext}$ is not provided, eq. (8) can be modified so that we can solve for $\bm{\ddot{q}}_{pred}$ expressed as

\begin{align}
\begin{split}
&\bm{0} = \sum_{k=1}^{p}c_k\left(\mathbf{M}_{k}\bm{\ddot{q}} + \mathbf{N}_{k}\bm{\dot{q}} - \mathbf{O}_{k} \right),\\
&-\left(\sum_{k=1}^{p}c_k\mathbf{M}_{k}\right)\bm{\ddot{q}} = \sum_{k=1}^{p}c_k\left(\mathbf{N}_{k}\bm{\dot{q}} - \mathbf{O}_{k} \right),\\
&\bm{\ddot{q}}_{pred} =  \left(-\sum_{k=1}^{p}c_k\mathbf{M}_{k}\right)^{-1} \sum_{k=1}^{p}c_k\left(\mathbf{N}_{k}\bm{\dot{q}} - \mathbf{O}_{k} \right) ,
\end{split}
\end{align}
where $\bm{\ddot{q}}_{pred}$ represents the predicted value of acceleration $\bm{\ddot{q}}$ and $\left(\cdot\right)^{-1}$ represents matrix inverse. In practice, we use Moore-Penrose pseudo inverse to calculate eq. (10) to avoid numerical instability. We define the following cost function to learn the Lagrangian of the system
\begin{align}
J(\bm{c}) = \| \bm{\ddot{q}} - \bm{\ddot{q}}_{pred}(\bm{c}) \|_{2}^{2}.
\end{align}

Due to the inverse operation, equation (11) is non-convex with respect to variable $\bm{c}$, making the optimization process not always converge to the global minimum. We empirically found that with a library of more than 20 candidate functions, the learning process will hardly converge, and the value of the cost function rarely touches a single-digit value even after a long period of iterations. Therefore, we only use this case only when no prior knowledge is available and the system is not complex such as a single pendulum. For more complex systems, such as a multi-DOF system, it is preferable that the external input $\bm{\tau}_{ext}$ is provided. In the case where no external input is provided, prior Lagrangian knowledge of a simpler system that forms a constituent of the larger system can be utilized to boost the learning process which will be explained in the next section. 

\begin{figure*}[t!]
  \vspace{0.2in}
  \centering
  \includegraphics[width=0.95\textwidth]{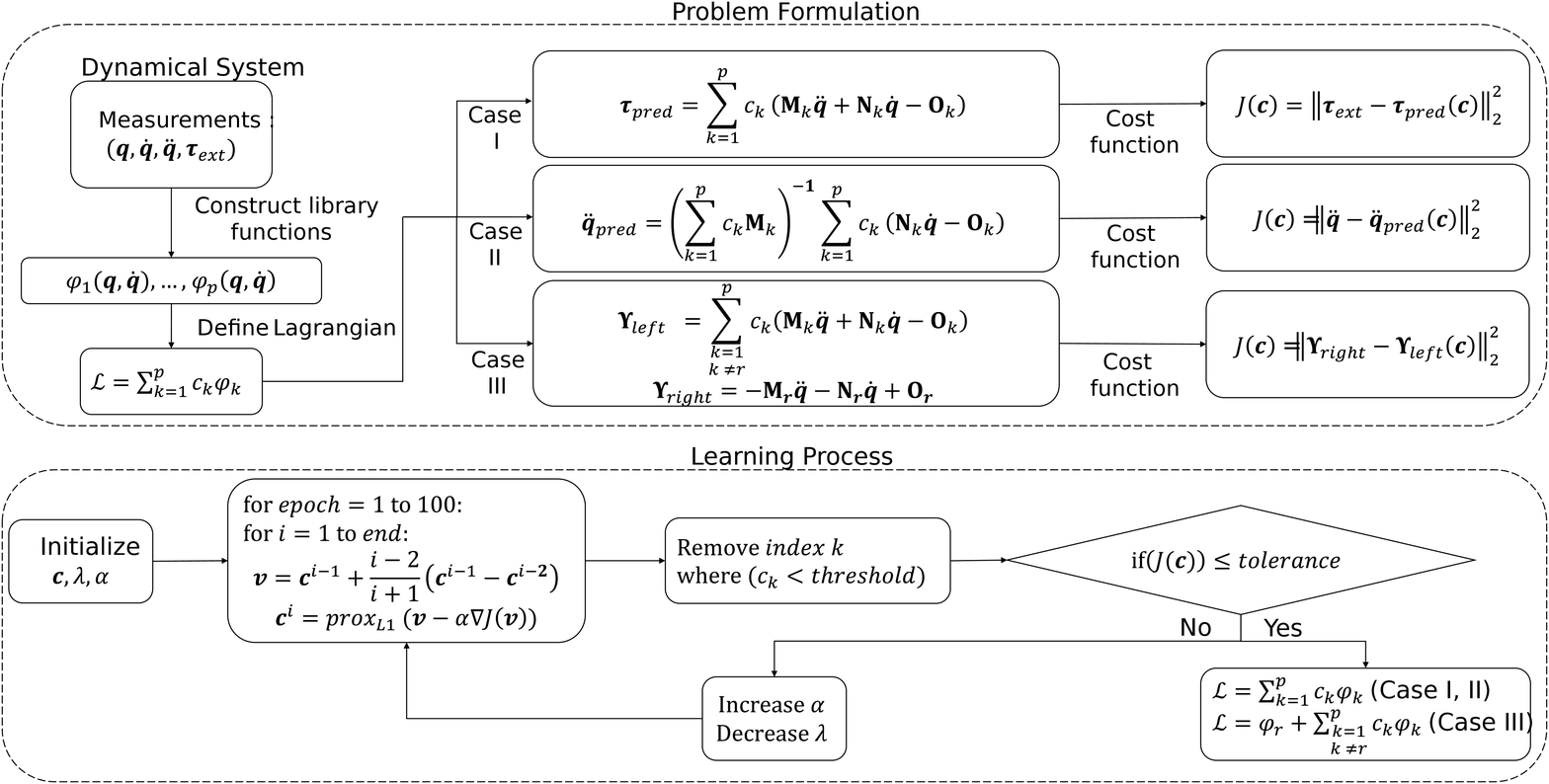}
  \caption{Block diagram of the proposed method (xL-SINDy). Depending on the case of the problem, a different cost function is constructed. Once the cost function is defined, the cost function is minimized by using the proximal gradient descent method.}
  \label{overview}
\end{figure*}

\subsubsection{With Prior Knowledge}

For multi-DOF non-relativistic systems, the Lagrangian can be described as 
$\mathcal{L} = \sum_i T_i - \sum_i V_i =   \sum_i \left( T_i - V_i \right)$,
where $T_i$ and $V_i$ are the kinetic energy and potential energy of each constituent of the system. Since the total Lagrangian of the system is the sum of the Lagrangian of its constituents, it is reasonable to assume that the nonlinear terms that appear in each constituent will also appear in the total Lagrangian of the system \cite{Hoang2020}. 

Given prior knowledge of a constituent of the system, we pick one out of several terms that appear in the total Lagrangian of the system and label them as $\phi_{r}(\bm{q},\bm{\dot{q}})$. The Lagrangian of a system is not unique; many forms of Lagrangians can satisfy Euler-Lagrange's equation for a particular system. For example, $\mathcal{L}' = k\mathcal{L}$, where $k$ is a constant, still satisfies the Euler-Lagrange's equation. By multiplying equation (1) with $k=\frac{1}{c_r}$, eq. (1) can be modified as

\begin{align}
&\mathcal{L} = \phi_r(\bm{q},\bm{\dot{q}}) + \sum\limits_{\substack{k=1 \\ k\neq r}}^{p} c'_k\phi_{k}(\bm{q},\bm{\dot{q}}),
\end{align}
where $c'_k = \frac{c_k}{c_r}$. Now, the variable $c'_k $ becomes the coefficients that we are interested to solve.  We can simply redefine $c_k := c'_k$ for notation simplicity. The Euler-Lagrange equation of the system can be expressed as
\begin{align}
\begin{split}
&\frac{d}{dt}\sum\limits_{\substack{k=1 \\ k\neq r}}^{p}c_k \nabla_{\bm{\dot{q}}}\phi_{k} - \sum\limits_{\substack{k=1 \\ k\neq r}}^{p}c_k \nabla_{\bm{q}}\phi_{k}\\ 
&= -\frac{d}{dt} \nabla_{\bm{\dot{q}}} \phi_r + \nabla_{\bm{q}} \phi_r.
\end{split}
\end{align}
We define the following notation

\begin{align}
\begin{split}
\bm{\Upsilon}_{left} &= \frac{d}{dt}\sum\limits_{\substack{k=1 \\ k\neq r}}^{p}c_k \nabla_{\bm{\dot{q}}}\phi_{k} - \sum\limits_{\substack{k=1 \\ k\neq r}}^{p}c_k \nabla_{\bm{q}}\phi_{k} \\
&=\sum\limits_{\substack{k=1 \\ k\neq r}}^{p}c_k\left(\mathbf{M}_{k}\bm{\ddot{q}} + \mathbf{N}_{k}\bm{\dot{q}} - \mathbf{O}_{k} \right),
\end{split}
\end{align}

\begin{align}
\begin{split}
\bm{\Upsilon}_{right} &= -\frac{d}{dt} \nabla_{\bm{\dot{q}}} \phi_r + \nabla_{\bm{q}} \phi_r \\
&= -\mathbf{M}_{r}\bm{\ddot{q}} - \mathbf{N}_{r}\bm{\dot{q}} + \mathbf{O}_{r},
\end{split}
\end{align}
where $\bm{\Upsilon}_{left}$ and $\bm{\Upsilon}_{right}$ represent the left hand side (LHS) and the right hand side (RHS) of eq. (13). By minimizing the following cost function

\begin{align}
J(\bm{c}) =\| \bm{\Upsilon}_{right} - \bm{\Upsilon}_{left}(\bm{c}) \|_{2}^{2},
\end{align}
it is possible to obtain the true Lagrangian of the system. We usually have more than one option  of $\phi_r$ to construct $\bm{\Upsilon}_{right}$. In practice, we have to test all of them one by one and choose the one that yields the best model.

\subsection{Learning Lagrangian}\label{subsec22}
The proposed learning method to obtain the Lagrangian is summarized in Fig. \ref{overview}. We start with the problem formulation as described in the previous section. Given a dynamical system, we gather times series data  $\{t_i,\bm{q}(t_{i}), \bm{\dot{q}}(t_{i}), \bm{\ddot{q})}(t_{i}, \bm{\tau}_{ext}(t_{i})\}_{i=1}^{N}$ from several initial conditions. We then proceed to construct a library of candidate functions. 

In general, the larger the library of the candidate functions, the more difficult the optimization problem becomes. It is especially true when several candidate functions can behave in a similar manner, such as in the trigonometric family functions. It is important to carefully construct a sufficient library but not too large so that the optimization problem is still tractable. It is also better, however, not to include trivial terms that satisfy Euler-Lagrange's equation regardless of trajectories such as $\mathcal{L} = \bm{q}^n\bm{\dot{q}}$. Technically, even though trivial terms will not affect the behavior of the systems, it is better to remove them so that we can reduce unnecessary complexity in our library. Depending on the case of the problem, a different cost function should be defined.

As mentioned previously, we believe that the correct solution is sparse where the majority of the coefficients are zero. Therefore, we add the L1 regularization term to the cost function for sparsity constraint \cite{lasso} expressed as
\begin{align}
J'(\bm{c}) = J(\bm{c}) + \lambda\|\bm{c}\|_1, 
\end{align}
where $\lambda$ is the sparsity promoting parameter that we have to carefully tune. In this work, we use the accelerated proximal gradient descent method \cite{FISTA} to minimize the composite cost function defined above. Given an initial point $\bm{c}^0$, the update step of proximal gradient descent is defined as 
\begin{align}
&\bm{v} = \bm{c}^{i-1} + \frac{i-2}{i+1}\left(\bm{c}^{i-1} - \bm{c}^{i-2}\right),\\
&\bm{c}^{i} = prox_{L1}\left(\bm{v} - \alpha\nabla J(\bm{v}) \right),
\end{align}
where ${c}^{i}$ is the coefficient $\bm{c}$ at iteration $i$, $\alpha$ is the learning rate, and $prox_{L1}(\cdot)$ is the proximal operator for L1 norm. The L1 norm penalty term is a separable sum of the component of its input, and a proximal operator is used to minimize this term. The proximal operator for the L1 norm is well defined separately for each component of the input and expressed as follows,
\begin{align}
&[prox_{L1}(\bm{\beta})]_{k} = \text{sign}(\beta_k)\max(|\beta_k|-\lambda,0),
\end{align}
where $k$ is the $k^{th}$ entry of the input vector $\bm{\beta}$. As for case III, if we know other terms that appear in the Lagrangian but are not used to construct $\bm{\Upsilon}_{right}$, we don't put a penalty on these terms by not applying the proximal operator in eq. (20) for index $k$ corresponding with these terms. 

We proceed to initialize the value of the coefficient $\bm{c}$, the learning rate $\alpha$,  and the L1 norm penalty parameter $\lambda$. The learning process is done in several stages, with 100 epochs and batch size equal to 128 for each stage, until the cost function reaches the defined tolerance value as shown in Fig. \ref{overview}. The tolerance value is ideally at $10^{-3}$. However, converging to this value might not be possible in the presence of noise, and we have to relax the tolerance value; otherwise, the algorithm will never stop.  In the beginning, the number of candidate functions in the library is usually large, and we want to eliminate non-relevant candidate functions as much as possible during the first learning stage. Therefore, we initially set the value of $\lambda$ to be quite high, which is between 1 and 5. 

\begin{figure}[b!]
\vspace{0.2in}
	\centering
	\includegraphics[scale=0.08]{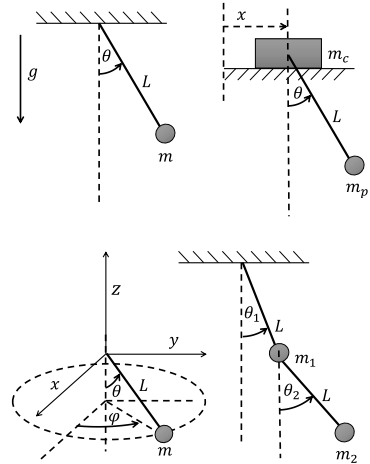}
	\caption{Dynamical systems used to verify xL-SINDy. From upper left to bottom right: A single pendulum, a cart pendulum, a spherical pendulum, and a double pendulum. For all systems, the length of the rod is $L = 1.0$ m,  the mass of all pendulums are $m = m_p = m_1 = m_2 = 1.0$ kg, and the gravitational acceleration is $g = 9.81$ m/s$^2$. For the cart-pendulum, the mass of the cart is $m_c = 0.5$ kg. }
	\label{systems}
\end{figure}

It is important to note that every candidate function may have different magnitude scales. The L1 norm penalty penalizes all terms equally regardless of the magnitude scale, resulting in candidate functions with smaller magnitude scales being penalized more. It may or may not be necessary to do scaling in the first learning stage, where the value of $\lambda$ is high, by multiplying each candidate function with scaling term $s_k$ in eq. (1) for cases I and II, or eq. (12) for case III depending on the differences of magnitude scale between each candidate function. The learning rate is also an important hyper-parameter, especially during the first learning stage. We found that a high learning rate in the initial stage can cause the relevant terms to be penalized, preventing the model from obtaining the true Lagrangian of the system. During the initial stage, we set the learning rate $\alpha <= 10^{-5}$.

At the end of every learning stage, we perform hard-thresholding by removing index $k$ from eq. (1) or (12), where the value of $c_k < threshold$. This step effectively reduces the number of candidate functions considered in the learning process, making the convergence much faster than if hard-thresholding were not performed. We then check whether the cost function has reached the tolerance or not. If it is the latter, we proceed to the next learning stage. With fewer candidate functions after the previous hard-thresholding process, we can decrease the value of $\lambda$ and increment the learning rate $\alpha$ to speed up the learning process. The rate of how we increment $\alpha$ and decrement $\lambda$ also matters. If we increment $\alpha$ too fast, the optimization process can even take longer to converge, and if we decrement $\lambda$ too fast, we might not be able to get rid of all of the non-relevant terms. A tuning process is needed to find the most appropriate rate to increase $\alpha$ and decrese $\lambda$. This step is repeated over and over again until the cost function reaches the tolerance value. Normally, the learning process will take around 3 or 4 stages until the tolerance value is reached. Once the tolerance value is reached, we compute the value of the coefficient to the eq. (1) for cases I and II, or (12) for case III, and we obtain the analytical form of Lagrangian of the system. 

\begin{table*}[tbp]
\vspace{0.2in}
\caption{Extracted Lagrangian from simulation data with various noise levels when external inputs are provided. All results are obtained with the computation described in case I.}
\begin{center}

\begin{tabular}{c c c c c}
\hline
\textbf{Noise Magnitude} & \textbf{\textit{Single Pendulum}}& \textbf{\textit{Cart Pendulum}}& \textbf{\textit{Double pendulum}}& \textbf{\textit{Spherical Pendulum}} \\
\hline

True Model & 
$
\begin{aligned}
&0.500\dot{\theta}^2 + 9.810\cos{\theta}
\end{aligned}
$ & $
\begin{aligned}
&0.250\dot{\theta}^2 + 0.750\dot{x}^2\\ 
&{+}\:0.500\dot{x}\dot{\theta}\cos{\theta} \\
&{+}\: 4.905\cos{\theta}
\end{aligned}
$ & $
\begin{aligned}
&19.620\cos{\theta_1} + 9.810\cos{\theta_2} \\
&{+}\:1.000\dot{\theta_1}\dot{\theta_2}\cos{\theta_1}\cos{\theta_2}\\
&{+}\:1.000\dot{\theta_1}\dot{\theta_2}\sin{\theta_1}\sin{\theta_2}\\
&{+}\:1.000\dot{\theta_1}^2+0.500\dot{\theta_2}^2
\end{aligned}
$ & $
\begin{aligned}
&0.500{\dot{\phi}^2\sin^2{\theta}}+0.500{\dot{\theta}^2}\\&{+}\:9.810\cos{\theta}
\end{aligned}
$ \\
\hline

$\sigma = 0$ & 
$
\begin{aligned}
&0.5\dot{\theta}^2 + 9.78\cos{\theta}
\end{aligned}
$ & $
\begin{aligned}
&0.25\dot{\theta}^2 + 0.75\dot{x}^2\\ 
&{+}\:0.5\dot{x}\dot{\theta}\cos{\theta} \\
&{+}\:4.89\cos{\theta}
\end{aligned}
$ & $
\begin{aligned}
&19.45\cos{\theta_1} + 9.72\cos{\theta_2} \\
&{+}\:0.99\dot{\theta_1}\dot{\theta_2}\cos{\theta_1}\cos{\theta_2}\\
&{+}\:0.99\dot{\theta_1}\dot{\theta_2}\sin{\theta_1}\sin{\theta_2}\\
&{+}\:0.99\dot{\theta_1}^2+0.5\dot{\theta_2}^2
\end{aligned}
$ & $
\begin{aligned}
&0.5{\dot{\phi}^2\sin^2{\theta}}+0.5{\dot{\theta}^2}\\&{+}\:9.76\cos{\theta}
\end{aligned}
$ \\
\hline

$\sigma = 10^{-3}$ & 
$
\begin{aligned}
&0.5\dot{\theta}^2 + 9.78\cos{\theta}
\end{aligned}
$ & $
\begin{aligned}
&0.25\dot{\theta}^2 + 0.75\dot{x}^2\\ 
&{+}\:0.5\dot{x}\dot{\theta}\cos{\theta}\\
&{+}\:4.88\cos{\theta}
\end{aligned}
$ & $
\begin{aligned}
&19.31\cos{\theta_1} + 9.65\cos{\theta_2} \\
&{+}\:0.99\dot{\theta_1}\dot{\theta_2}\cos{\theta_1}\cos{\theta_2}\\
&{+}\:0.99\dot{\theta_1}\dot{\theta_2}\sin{\theta_1}\sin{\theta_2}\\
&{+}\:0.99\dot{\theta_1}^2+0.49\dot{\theta_2}^2
\end{aligned}
$ & $
\begin{aligned}
&0.5{\dot{\phi}^2\sin^2{\theta}}+0.5{\dot{\theta}^2}\\&{+}\:9.8\cos{\theta}
\end{aligned}
$ \\
\hline

$\sigma = 2\times10^{-2}$ & 
$
\begin{aligned}
&0.49\dot{\theta}^2 + 9.66\cos{\theta}
\end{aligned}
$ & $
\begin{aligned}
&0.24\dot{\theta}^2 + 0.72\dot{x}^2\\ 
&{+}\:0.48\dot{x}\dot{\theta}\cos{\theta}\\
&{+}\:4.69\cos{\theta}
\end{aligned}
$ & $
\begin{aligned}
&17.52\cos{\theta_1} + 8.76\cos{\theta_2} \\
&{+}\:0.99\dot{\theta_1}\dot{\theta_2}\cos{\theta_1}\cos{\theta_2}\\
&{+}\:0.89\dot{\theta_1}\dot{\theta_2}\sin{\theta_1}\sin{\theta_2}\\
&{+}\:0.89\dot{\theta_1}^2+0.45\dot{\theta_2}^2
\end{aligned}
$ & $
\begin{aligned}
&0.31{\dot{\phi}^2\sin^2{\theta}}+0.31{\dot{\theta}^2}\\&{+}\:6.09\cos{\theta}
\end{aligned}
$ \\
\hline

$\sigma = 6\times10^{-2}$ & 
$
\begin{aligned}
&0.44\dot{\theta}^2 + 8.73\cos{\theta}
\end{aligned}
$ & $
\begin{aligned}
&0.17\dot{\theta}^2 + 0.53\dot{x}^2\\ 
&{+}\:0.36\dot{x}\dot{\theta}\cos{\theta}\\
&{+}\:3.32\cos{\theta} 
\end{aligned}
$ & $
\begin{aligned}
&10.72\cos{\theta_1} + 5.42\cos{\theta_2} \\
&{+}\:0.55\dot{\theta_1}\dot{\theta_2}\cos{\theta_1}\cos{\theta_2}\\
&{+}\:0.55\dot{\theta_1}\dot{\theta_2}\sin{\theta_1}\sin{\theta_2}\\
&{+}\:0.55\dot{\theta_1}^2+0.27\dot{\theta_2}^2
\end{aligned}
$ & $
\begin{aligned}
&-0.66{\dot{\phi}^2\sin^2{\theta}}-0.06{\dot{\theta}^2}\\&{-}\:1.94\cos{\theta}\\ &\highlight{{+}\:0.08\dot{\theta}\dot{\phi}}\\
&\highlight{{-}\:0.03\phi^2}^*
\end{aligned}
$ \\
\hline

$\sigma = 10^{-1}$ & 
$
\begin{aligned}
&0.36\dot{\theta}^2 + 7.19\cos{\theta} \\
\end{aligned}
$ & $
\begin{aligned}
&0.11\dot{\theta}^2 + 0.36\dot{x}^2 + \\ 
&0.24\dot{x}\dot{\theta}\cos{\theta} \\
&{+}\:2.03\cos{\theta} \\
\end{aligned}
$ & $
\begin{aligned}
&5.18\cos{\theta_1} + 2.66\cos{\theta_2} \\
&{+}\:0.26\dot{\theta_1}\dot{\theta_2}\cos{\theta_1}\cos{\theta_2}\\
&{+}\:0.26\dot{\theta_1}\dot{\theta_2}\sin{\theta_1}\sin{\theta_2}\\
&{+}\:0.26\dot{\theta_1}^2+0.13\dot{\theta_2}^2
\end{aligned}
$ & $
\begin{aligned}
&0.04{\dot{\phi}^2\sin^2{\theta}}+0.04{\dot{\theta}^2}\\&{+}\:1.59\cos{\theta}\\
&\highlight{{-}\:0.1\dot{\theta}\phi\sin{\theta}} \\
&\highlight{{-}\:0.22\phi^2}^* 
\end{aligned}
$ \\
\hline
\multicolumn{4}{l}{$^*$Terms highlighted with red color are extra terms that are not supposed to be included in the Lagrangian.} \\
\multicolumn{4}{l}{ All numbers are rounded to 3 decimal places.}
\end{tabular}
\end{center}
\label{result-table-active}
\end{table*}

\begin{table*}[tbp]
\vspace{0.2in}
\caption{Extracted Lagrangian from simulation data with various noise levels when no external input is provided.  For the single pendulum, the result is obtained with a computation described by case II, while the rest are obtained with a computation described by case III with the knowledge of the Lagrangian of a single pendulum.}
\begin{center}
\begin{tabular}{c c c c c}
\hline
\textbf{Noise Magnitude} & \textbf{\textit{Single Pendulum}}& \textbf{\textit{Cart Pendulum}}& \textbf{\textit{Double pendulum}}& \textbf{\textit{Spherical Pendulum}} \\
\hline

True Model & 
$
\begin{aligned}
&0.500\dot{\theta}^2 + 9.810\cos{\theta}
\end{aligned}
$ & $
\begin{aligned}
&0.250\dot{\theta}^2 + 0.750\dot{x}^2\\ 
&{+}\:0.500\dot{x}\dot{\theta}\cos{\theta} \\
&{+}\: 4.905\cos{\theta}
\end{aligned}
$ & $
\begin{aligned}
&19.620\cos{\theta_1} + 9.810\cos{\theta_2} \\
&{+}\:1.000\dot{\theta_1}\dot{\theta_2}\cos{\theta_1}\cos{\theta_2}\\
&{+}\:1.000\dot{\theta_1}\dot{\theta_2}\sin{\theta_1}\sin{\theta_2}\\
&{+}\:1.000\dot{\theta_1}^2+0.500\dot{\theta_2}^2
\end{aligned}
$ & $
\begin{aligned}
&0.500{\dot{\phi}^2\sin^2{\theta}}+0.500{\dot{\theta}^2}\\&{+}\:9.810\cos{\theta}
\end{aligned}
$ \\
\hline

$\sigma = 0$ & 
$
\begin{aligned}
&0.295\dot{\theta}^2 + 5.797\cos{\theta}
\end{aligned}
$ & $
\begin{aligned}
&1.000\dot{\theta}^2 + 2.975\dot{x}^2\\ 
&{+}\:1.984\dot{x}\dot{\theta}\cos{\theta} \\
&{+}\:19.755\cos{\theta}
\end{aligned}
$ & $
\begin{aligned}
&19.620\cos{\theta_1} + 9.750\cos{\theta_2} \\
&{+}\:1.000\dot{\theta_1}\dot{\theta_2}\cos{\theta_1}\cos{\theta_2}\\
&{+}\:0.999\dot{\theta_1}\dot{\theta_2}\sin{\theta_1}\sin{\theta_2}\\
&{+}\:1.000\dot{\theta_1}^2+0.499\dot{\theta_2}^2
\end{aligned}
$ & $
\begin{aligned}
&1.000{\dot{\phi}^2\sin^2{\theta}}+1.000{\dot{\theta}^2}\\&{+}\:19.630\cos{\theta}
\end{aligned}
$ \\
\hline

$\sigma = 10^{-3}$ & 
$
\begin{aligned}
&0.268\dot{\theta}^2 + 5.252\cos{\theta}
\end{aligned}
$ & $
\begin{aligned}
&1.000\dot{\theta}^2 + 2.975\dot{x}^2\\ 
&{+}\:1.984\dot{x}\dot{\theta}\cos{\theta}\\
&{+}\:19.756\cos{\theta}
\end{aligned}
$ & $
\begin{aligned}
&19.508\cos{\theta_1} + 9.755\cos{\theta_2} \\
&{+}\:1.000\dot{\theta_1}\dot{\theta_2}\cos{\theta_1}\cos{\theta_2}\\
&{+}\:0.999\dot{\theta_1}\dot{\theta_2}\sin{\theta_1}\sin{\theta_2}\\
&{+}\:1.000\dot{\theta_1}^2+0.499\dot{\theta_2}^2
\end{aligned}
$ & $
\begin{aligned}
&1.000{\dot{\phi}^2\sin^2{\theta}}+1.000{\dot{\theta}^2}\\&{+}\:19.630\cos{\theta}
\end{aligned}
$ \\
\hline

$\sigma = 2\times10^{-2}$ & 
$
\begin{aligned}
&0.334\dot{\theta}^2 + 6.540\cos{\theta}
\end{aligned}
$ & $
\begin{aligned}
&1.000\dot{\theta}^2 + 2.993\dot{x}^2\\ 
&{+}\:1.994\dot{x}\dot{\theta}\cos{\theta}\\
&{+}\:19.534\cos{\theta}
\end{aligned}
$ & $
\begin{aligned}
&19.545\cos{\theta_1} + 9.770\cos{\theta_2} \\
&{+}\:1.000\dot{\theta_1}\dot{\theta_2}\cos{\theta_1}\cos{\theta_2}\\
&{+}\:0.999\dot{\theta_1}\dot{\theta_2}\sin{\theta_1}\sin{\theta_2}\\
&{+}\:1.000\dot{\theta_1}^2+0.499\dot{\theta_2}^2
\end{aligned}
$ & $
\begin{aligned}
&1.000{\dot{\phi}^2\sin^2{\theta}}+1.000{\dot{\theta}^2}\\&{+}\:19.600\cos{\theta}
\end{aligned}
$ \\
\hline

$\sigma = 6\times10^{-2}$ & 
$
\begin{aligned}
&0.557\dot{\theta}^2 + 10.938\cos{\theta}
\end{aligned}
$ & $
\begin{aligned}
&1.000\dot{\theta}^2 + 1.696\dot{x}^2\\ 
&{+}\:1.136\dot{x}\dot{\theta}\cos{\theta}\\
&{+}\:18.082\cos{\theta} \\ &\highlight{{-}\:0.121\dot{x}^2\cos{\theta}} \\
&\highlight{{+}\:1.463\cos^3{\theta}}^*
\end{aligned}
$ & $
\begin{aligned}
&19.541\cos{\theta_1} + 9.753\cos{\theta_2} \\
&{+}\:0.999\dot{\theta_1}\dot{\theta_2}\cos{\theta_1}\cos{\theta_2}\\
&{+}\:0.999\dot{\theta_1}\dot{\theta_2}\sin{\theta_1}\sin{\theta_2}\\
&{+}\:1.000\dot{\theta_1}^2+0.496\dot{\theta_2}^2
\end{aligned}
$ & $
\begin{aligned}
&0.130{\dot{\phi}^2\sin^2{\theta}}+1.000{\dot{\theta}^2}\\&{+}\:2.350\cos{\theta}\\ &\highlight{{-}\:0.790\dot{\theta}^2\sin{\theta}}\\
&\highlight{{-}\:0.430\dot{\theta}^2\cos{\theta}}^*
\end{aligned}
$ \\
\hline

$\sigma = 10^{-1}$ & 
$
\begin{aligned}
&0.085\dot{\theta}^2 + 1.540\cos{\theta} \\
&\highlight{{-}\:0.129\dot{\theta}\sin{\theta}} \\
&\highlight{{+}\:0.551\sin^2{\theta}}\\
&\highlight{{-}\:0.019\theta^2}^*
\end{aligned}
$ & $
\begin{aligned}
&1.000\dot{\theta}^2 + 1.562\dot{x}^2 + \\ 
&1.050\dot{x}\dot{\theta}\cos{\theta} \\
&{+}\:19.504\cos{\theta} \\
&\highlight{{-}\:0.143\dot{\theta}^2\cos{\theta}}^*
\end{aligned}
$ & $
\begin{aligned}
&19.381\cos{\theta_1} + 9.679\cos{\theta_2} \\
&{+}\:0.998\dot{\theta_1}\dot{\theta_2}\cos{\theta_1}\cos{\theta_2}\\
&{+}\:0.992\dot{\theta_1}\dot{\theta_2}\sin{\theta_1}\sin{\theta_2}\\
&{+}\:1.000\dot{\theta_1}^2+0.495\dot{\theta_2}^2
\end{aligned}
$ & $
\begin{aligned}
&-0.2{\dot{\phi}^2\sin^2{\theta}}+1.000{\dot{\theta}^2}\\&{+}\:5.12\cos{\theta}\\
&\highlight{{-}\:1.100\dot{\theta}^2\sin{\theta}} \\
&\highlight{{-}\:0.560\dot{\theta}^2\cos{\theta}} \\
&\highlight{{-}\:0.055\dot{\phi}^2\sin{2\theta}}^* 
\end{aligned}
$ \\
\hline
\multicolumn{4}{l}{$^*$Terms highlighted with red color are extra terms that are not supposed to be included in the Lagrangian.} \\
\multicolumn{4}{l}{ All numbers are rounded to 3 decimal places.}
\end{tabular}
\end{center}
\label{result-table-passive}
\end{table*}

\subsection{Dynamical Systems and Experiments}
We evaluated xL-SINDy with four ideal dynamical systems as shown in Fig. \ref{systems}. In this work, at first, we test xL-SINDy when the dynamical systems are excited by external inputs $\bm{\tau}_{ext} = \bm{f}(\sin{\omega t}, \cos{\omega t}) $, where $\omega$ is a random frequency. The computation used to learn the model is described by the computation of case I. 

We also test the case of passive systems where no external input $\bm{\tau}_{ext}$ is provided. For the case of a single pendulum, we assume that no prior knowledge is available. Thus, we use the computation described in case II. For the cart pendulum, double pendulum, and spherical pendulum, these systems either have a single pendulum as one of their constituents or a more complex version of a single pendulum. Thus, assuming that we already obtained the Lagrangian of a single pendulum, we can use the computation described in case III to bootstrap the learning process.

For each system, we collect training data by performing simulation with 100 initial conditions for a period of 5s each and 100 Hz of measurement frequency. After obtaining the analytical form of the Lagrangian, we create a validation data set to test the obtained model by calculating the predicted states for the accuracy evaluation. We compute the Euler-Lagrange's equation with the obtained model, retrieve the differential equation of the system, and integrate the equations to compare them with the actual validation data. We also tested our method with training data that are corrupted by zero-mean white Gaussian noise $\mathcal{N}(0, \sigma)$ on different scale magnitude in the range of $10^{-8} <= \sigma <= 10^{-1}$. Finally, we also compare the performance of xL-SINDy on several passive dynamical systems with noisy training data against SINDy-PI \cite{Kaheman2020}. 

\begin{figure*}[t!]
\vspace{0.2in}
\centering
  \includegraphics[width=1\textwidth]{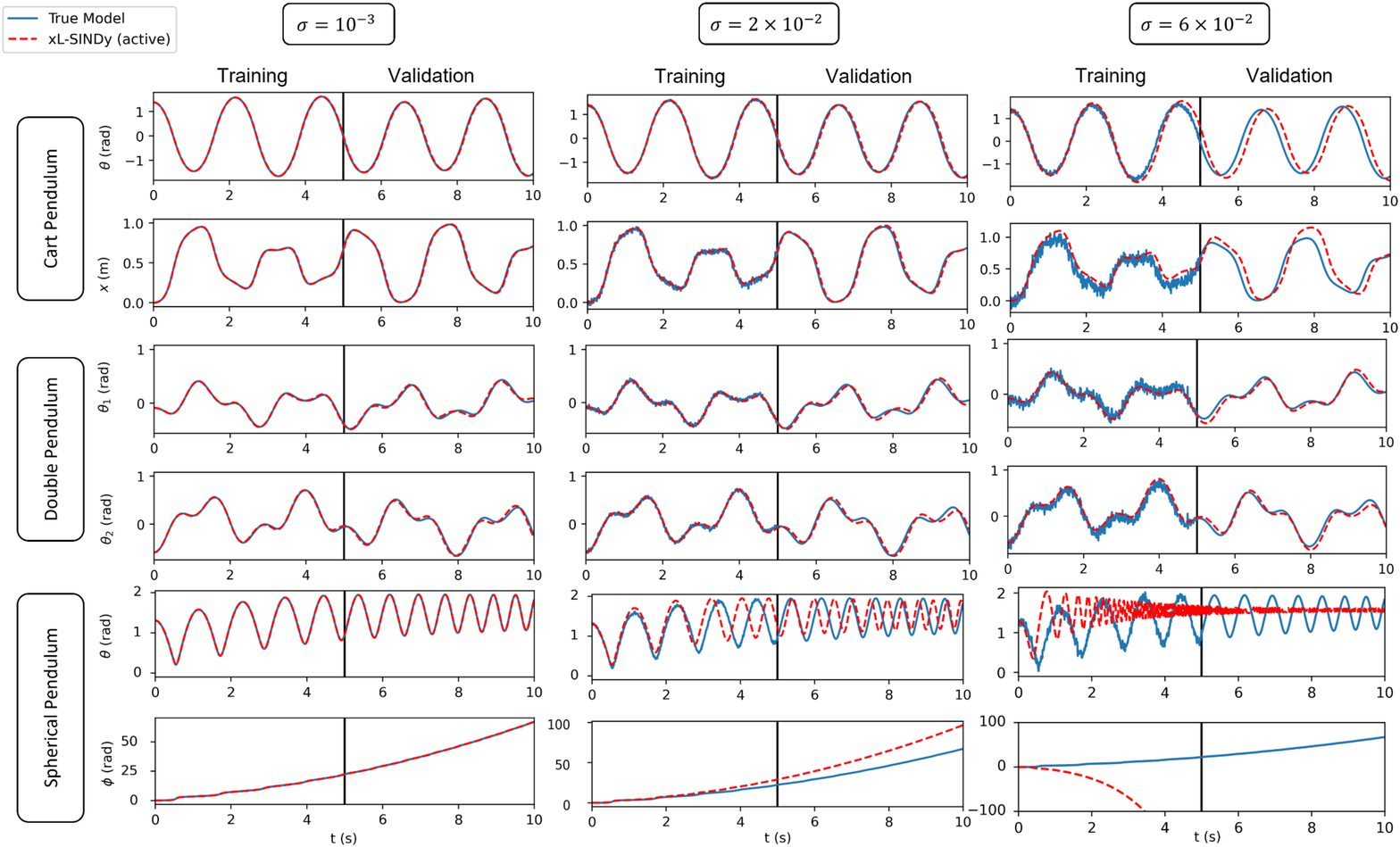}
  \caption{Comparison against true model when external excitation is provided. Training data consists of 100 initial conditions in a time period of 5 seconds each. Validation (extrapolation beyond the training data set) is conducted for 5 seconds afterward. The results shown are taken randomly from one of the initial conditions from the training data set for cart pendulum, double pendulum, and spherical pendulum. }
  \label{results-active}
\end{figure*}

\begin{figure*}[t!]
\vspace{0.2in}
\centering
  \includegraphics[width=1\textwidth]{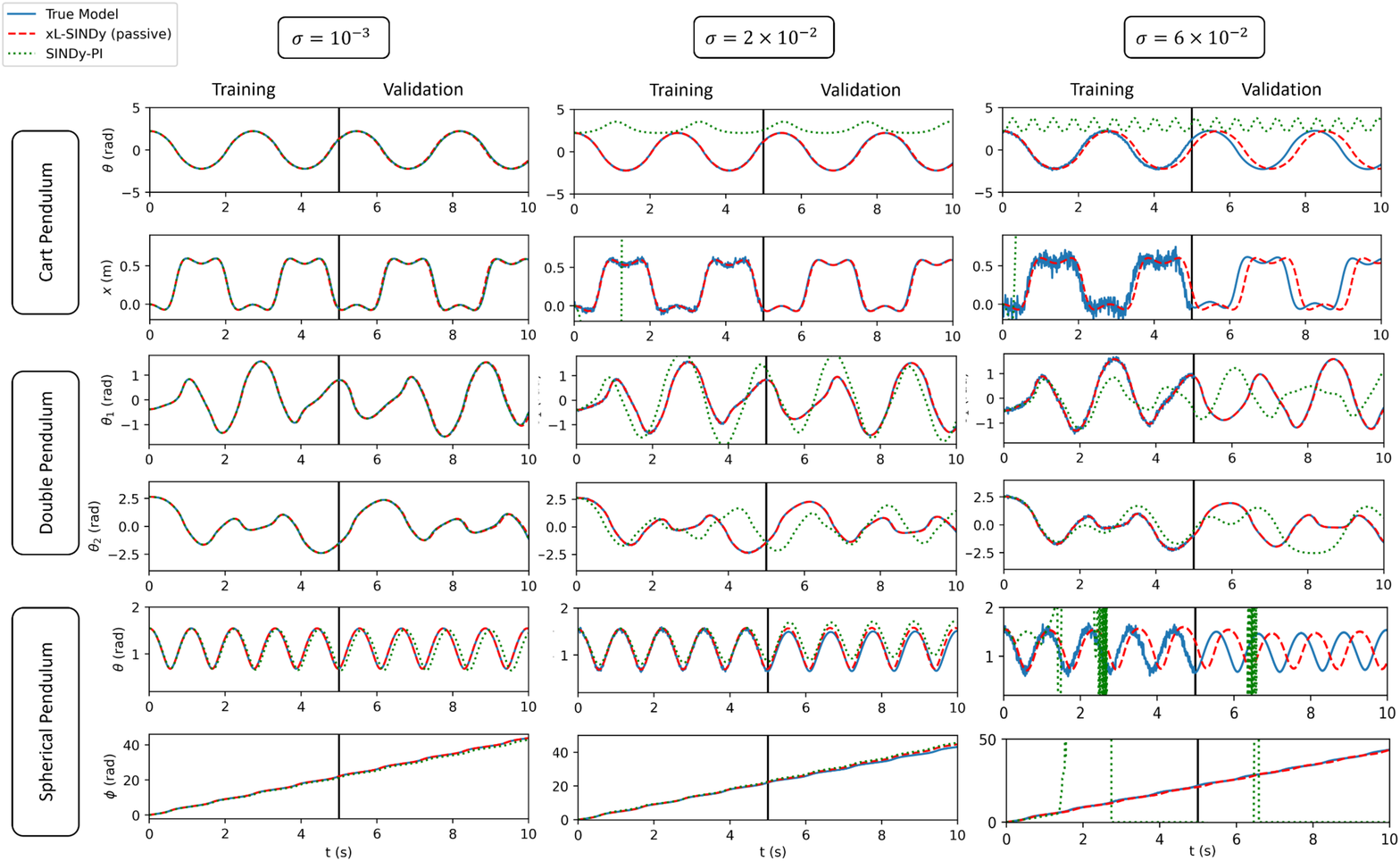}
  \caption{Simulation results against three different noise levels for three different models: the true model, model discovered by the proposed method, and model discovered by SINDy-PI. Training data consists of 100 initial conditions in a time period of 5 seconds each. Validation (extrapolation beyond the training data set) is conducted for 5 seconds afterward. The results shown are taken randomly from one of the initial conditions from the training data set for cart pendulum, double pendulum, and spherical pendulum. }
  \label{results-passive}
\end{figure*}

\section{Results}
In this section, we demonstrate the effectiveness of xL-SINDy with the aforementioned nonlinear dynamical systems against various noise levels.  The obtained Lagrangian for each system is summarized in Table. \ref{result-table-active} for the case of active systems, and in Table. \ref{result-table-passive} for the case of passive systems. The performance of xL-SINDy against the true model from our simulation experiments on a cart pendulum, a double pendulum, and a spherical pendulum is shown in Fig. \ref{results-active} in the case of active systems, and in Fig. \ref{results-passive} in the case of passive systems along with the comparison against SINDy-PI. From Table. \ref{result-table-active} and Table. \ref{result-table-passive}, it can be seen that xL-SINDy performs better to extract the correct structure if external inputs are provided. It includes fewer wrong additional terms in the model compared to when no external input is provided. 

As for the performance of xL-SINDy against SINDy-PI, in the second column of the plot in Fig. \ref{results-passive}, where the noise magnitude is $\sigma = 2\times10^{-2}$, we can observe that SINDy-PI has already started to deviate from the true models in all three dynamical systems.  At the same noise magnitude, xL-SINDy still predicts accurate models. It is also good to note that the model estimate of xL-SINDy is still reasonable even though wrong additional terms are included in the Lagrangian from the example of the cart pendulum under the noise magnitude of $\sigma = 6\times10^{-2}$. It indicates that the model estimate is potentially usable even when an incorrect Lagrangian structure is discovered. Our simulation results demonstrate that xL-SINDy has notably better prediction accuracy compared to SINDy-PI in the presence of higher noise magnitude in all three dynamical systems.   

\subsection{Single Pendulum}
The state of a single pendulum is described by $[\theta, \dot{\theta}]$, and the Lagrangian expression of a single pendulum is given by 
$
\mathcal{L} = \frac{1}{2}m\dot{\theta}^2 + mg\cos{\theta},
$
Substituting the parameter given in Fig. \ref{systems}, the true Lagrangian expression is shown in the second row and second column of Table. \ref{result-table-active}. To construct a library of candidate functions, we create a polynomial combination of $\{\theta, \dot{\theta}, \cos{\theta}, \sin{\theta}\}$ up to the second order while excluding trivial terms such as $\dot{\theta}$ and $\theta\dot{\theta}$ resulting in 12 candidate functions. Training data with initial conditions of $[-\pi < \theta < \pi,0]$ are created. 

The initial value of the hyperparameters are $\alpha = 10^{-5}$ and $\lambda = 0.1$. The cut-off threshold is $10^{-2}$ for the initial learning stage and $10^{-1}$ for the subsequent learning stages. In the subsequent learning stages, $\alpha$ is increased by a factor of 2, and $\lambda$ is decreased by a factor of 10. The training converged in three stages for noise magnitude $\sigma <= 10^{-3}$, and four stages for higher magnitude with a relaxed tolerance value. 

In the case of an active system, the correct Lagrangian structure, the ones without additional terms or missing terms compared to the true Lagrangian form, can be obtained in the presence of noise magnitude up to $\sigma = 1\times10^{-1}$. While
in the case of a passive system, the correct Lagrangian structure can be obtained
with noise magnitude up to $\sigma = 6\times10^{-2}$. Even though the coefficients obtained differ from the true model, the ratio of coefficients between the two terms is close compared to the true model.

\subsection{Cart Pendulum}
The state of the cart pendulum is represented as $[\theta, \dot{\theta}, x, \dot{x}]$, and the Lagrangian with numerical coefficients is shown in the second row and third column of Table. \ref{result-table-active} with parameters given by Fig. \ref{systems}. A library of candidate function with a polynomial combination of $\{\dot{\theta}, \cos{\theta}, \sin{\theta}, x, \dot{x}\}$ up to the third order is constructed, resulting in 55 candidate functions.We here exclude the term $\theta$ because this term does not appear in the Lagrangian of a single pendulum system. Training data with initial conditions of $[-\pi < \theta < \pi, 0, 0,0]$ are created.

The Lagrangian of a single pendulum contains $\dot{\theta}^2$ and $\cos{\theta}$. Hence, both terms will also appear in the Lagrangian of the cart pendulum. We tested both $\dot{\theta}^2$ and $\cos{\theta}$ to construct $\bm{\Upsilon}_{right}$ as described in eq. (15), and we found that the term  $\dot{\theta}^2$ gives better results. The initial value of the hyperparameters are $\alpha = 10^{-5}$ and $\lambda = 1$. The cut-off threshold, the increment of $\alpha$, and the decrement of $\lambda$ are the same as in the previous case. The training converged in three stages for noise magnitude $\sigma <= 2\times10^{-2}$, and four stages for higher magnitude with a relaxed tolerance value. 

From the table, xL-SINDy can recover the correct structure of the Lagrangian with noise magnitude up to $\sigma = 1\times10^{-1}$ with external input and noise magnitude up to $\sigma = 4\times10^{-2}$ without external input. In contrast, SINDy-PI can only recover the correct structure with noise magnitude up to $\sigma = 5\times10^{-3}$. Therefore, in the case of the cart pendulum, xL-SINDy is 8 times more robust than SINDy-PI in the presence of noise. On top of that, with a large magnitude of noises, SINDy-PI sometimes predicts a model which blows up as it is shown in the second column of Fig. \ref{results-passive}. This happens because it incorrectly predicts the denominator terms. Unlike SINDy-PI, xL-SINDy deals with the Lagrangian instead of the actual equation of motion. Since there is no denominator in the Lagrangian, xl-SINDy still gives reasonable predictions even though wrong additional terms are included in the Lagrangian.

\subsection{Double Pendulum}
Given the state of a double pendulum, $[\theta_1, \theta_2, \dot{\theta_1}, \dot{\theta_2}]$ and the system parameters in Fig. \ref{systems}, the expression of the Lagrangian with numerical coefficients is shown in the second row and fourth column of Table. \ref{result-table-active}. To build a library of candidate functions, we first separate the set of trigonometric terms $\{\cos{\theta_1},\sin{\theta_1},\sin{\theta_1},\sin{\theta_2}\}$, and the non trigonometric terms $\{\dot{\theta_1}, \dot{\theta_2}\}$. For each set, we create a polynomial combination up to the second order, resulting in 14 candidate functions and 5 candidate functions respectively. We then generate cross terms between the two sets creating 70 candidate functions, and we have in total 89 candidate functions in the library.  Training data are created with initial conditions of $[-\pi < \theta_1 < \pi,-\pi < \theta_2 < \pi,0,0]$.

Both constituents of the double pendulum are single pendulums. Hence, we have 4 options to construct $\bm{\Upsilon}_{right}$: $\dot{\theta_1}^2$, $\dot{\theta_2}^2$, $\cos{\theta_1}$, and $\cos{\theta_2}$. Both $\dot{\theta_1}^2$ and $\dot{\theta_2}^2$ yield equally good results. The results displayed in Table. \ref{result-table-active} and Table. \ref{result-table-passive} are the one with $\dot{\theta_1}^2$ used to construct $\bm{\Upsilon}_{right}$. The initial value of the hyperparameters are $\alpha = 5\times10^{-6}$ and $\lambda = 1$. The cut-off threshold, the increment of $\alpha$, and the decrement of $\lambda$ are the same as in previous cases. From our experiments, xL-SINDy can identify the correct structure with noise magnitude up to $\sigma = 10^{-1}$ in both active and passive cases, while SINDy-PI can extract the correct structure of equations of motions with noise magnitude up to $\sigma = 10^{-2}$. Hence, xL-SINDy is 10 times more robust against noise than SINDy-PI in this experiment. 

As it can be seen from the summary table that xL-SINDy is quite robust in the case of double pendulum compared to other dynamical systems.  One possible reason why xL-SINDy is more robust in the case of the double pendulum is due to the chaotic signal caused by the double pendulum. For every initial condition, the double pendulum will yield an entirely different signal path due to the inherently chaotic nature of the double pendulum creating rich training data. 

\subsection{Spherical Pendulum}
The state of the cart pendulum is represented as $[\theta, \phi, \dot{\theta}, \dot{\phi}]$, and the true Lagrangian expression is displayed in the second row and fifth column of Table. \ref{result-table-active}. As in the case of the double pendulum, we first separate the trigonometric terms $\{\cos{\theta},\sin{\theta}\}$ and the non-trigonometric terms $\{\dot{\theta},\phi,\dot{\phi}\}$, create polynomial combinations for both sets up to the second order and add cross terms between the two sets. In total, we have 59 candidate functions in our library. The training data are created with initial conditions of $[\pi/3 < \theta < \pi/2,0,0,\pi]$. We deliberately choose high value of $\theta$ and $\dot{\phi}$ as the initial conditions because the equation of motion contains $\frac{1}{\sin{\theta}}$ which could blow up for small value of $\theta$.

The spherical pendulum is a higher dimensional analog of a single pendulum in the first case. Therefore, we can think of the Lagrangian of a spherical pendulum as the sum of the Lagrangian of the pendulum in $\hat{\theta}$ direction and $\hat{\phi}$ direction. Since we already know the Lagrangian of a single pendulum in $\hat{\theta}$ direction, we can use $\dot{\theta}^2$ and $\cos{\theta}$ to construct $\bm{\Upsilon}_{right}$. The initial value of the hyperparameters are $\alpha = 1\times10^{-5}$ and $\lambda = 1$. The cut-off threshold, the increment of $\alpha$, and the decrement of $\lambda$ are the same as in all previous cases. In this experiment, xL-SINDy is robust only up to $\sigma = 2 \times 10^{-2}$ for both active and passive system cases. In contrast, SINDy-PI is only robust up to $\sigma = 1 x 10^{-3}$. Thus, xL-SINDy shows 20 times more robustness against noise than SINDy-PI. 

From the second column of Fig. \ref{results-active} and Fig. \ref{result-table-passive}, it can be seen that even though the correct structure can be obtained for the spherical pendulum at $\sigma = 2 \times 10^{-2}$, the performance of xL-SINDy in the case of an active system is worse. This is because in the case of the passive system, a Lagrangian multiplied by a constant is still a valid Lagrangian, thus as long as the ratio between each coefficient is the same as the true model, then the obtained model is also correct. However, in the case of the active system, since there is an external constrain, the Lagrangian is unique. So, even though the correct structure is obtained, if the coefficient does not closely match the true model, it will less accurate long-term prediction ability.

\begin{figure}[t!]
\vspace{0.2in}
\centering
  \includegraphics[scale = 0.5]{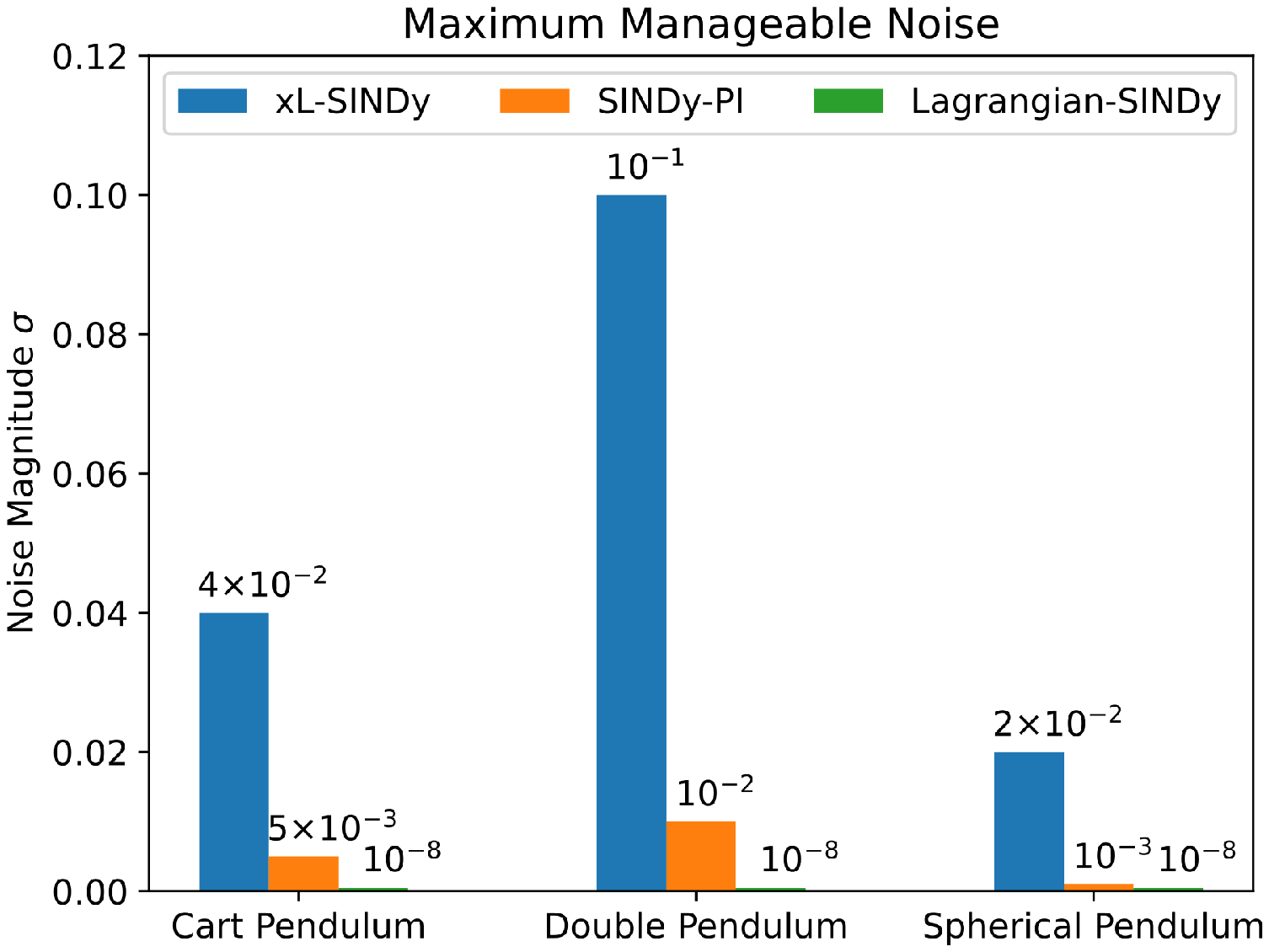}
  \caption{Comparison of the maximum level of manageable noise in training data before an incorrect model structure is discovered for three methods: xL-SINDy, SINDy-PI, Lagrangian-SINDy. xL-SINDy provides the most robust performance against noisy training data in all simulations of three different dynamical systems.}
  \label{noise}
\end{figure}

\begin{table}[t!]
\centering
\caption{Number of terms used in the library to obtain the model in different dynamical systems.}
\resizebox{\columnwidth}{!}{
\begin{tabular}{|l| c| c| c|}
\hline
\backslashbox{Method}{Dynamical\\Sytem} & \makecell{Cart\\Pendulum} & \makecell{Double\\Pendulum} & \makecell{Spehrical\\Pendulum}\\
\hline
\makecell{xL-SINDy} & 55 & 87 & 59\\
\hline
\makecell{SINDy-PI} & 90 & 40 & 49\\
\hline
\end{tabular}}
\label{table-num-library}
\end{table}

\section{Discussion}
From the results of simulations, it can be concluded that xL-SINDy is more robust to obtaining the correct Lagrangian structure if external input is provided to the system. However, with larger noise, even though xL-SINDy can obtain the correct Lagrangian structure, it does not guarantee to provide an accurate long-term prediction ability. This is because in the case where external input is provided, the Lagrangian of the system is unique, thus a mismatch coefficient will result in deviation in the long-term prediction. However, this will not be a problem if there is no external input as the Lagrangian is not unique. As long as the ratio between each coefficient is close to the true model, it can still give a good long-term prediction model. However, with the absence of external input, xL-SINDy is a little bit less robust to find the correct Lagrangian structure.

To compare xL-SINDy with other methods, we use passive cases as the baseline of xL-SINDy. The comparison of xL-SINDy, SINDy-PI, and Lagrangian-SINDy is summarized in Fig. \ref{noise}. In all three dynamical systems used as a comparison, xL-SINDy outperforms other methods in terms of noise robustness. Our experiment results demonstrate that xL-SINDy can overcome the challenge faced by Lagrangian-SINDy. xL-SINDy is capable of discovering the correct Lagrangian expression for idealized nonlinear dynamical systems in the presence of much higher noise magnitude. On top of that, xL-SINDy successfully extracts the Lagrangian in cases where Lagrangian-SINDy fails to do so, such as the non-actuated spherical pendulum \cite{Hoang2020}. The obtained coefficients may not be precisely the same as the true models, but the ratio between coefficients of each term is close to the true models.

From our experiments, while SINDy-PI is also robust against noise up to a certain magnitude, xL-SINDy is 8 to 20 times more robust against noise. SINDy-PI attempts to seek the expression of the dynamics which may contain rational functions. To do so, SINDy-PI reformulates the problem into implicit form, and it requires the library to include candidate functions of the states and the time derivative of the states of the systems. As a consequence, SINDy-PI also has to include $\bm{q}$, $\bm{\dot{q}}$, and $\bm{\ddot{q}}$ variables to build terms in the library. Unlike SINDy-PI which deals with the actual equation of motion, xL-SINDy deals with Lagrangian that only requires $\bm{q}$ and $\dot{q}$ variables to build the terms in the library. Hence, naturally, to obtain the same order of a family function (e.g. second order of polynomial functions), xL-SINDy would contain fewer terms in the library.

The second advantage of xL-SINDy over SINDy-PI is that xL-SINDy can have more coverage of various terms with fewer coefficient parameters in the actual equation of motion. This is due to the inherent nature of the Lagrangian formulation itself. For example, let's consider a Lagrangian that only contains one term given by
\begin{align}
\mathcal{L} = c_0\phi_0(\bm{q}, \bm{\dot{q}}).
\end{align}
 If we substitute this into Euler-Lagrange's formula, we will get the equation of motion that contains three different terms expressed as
\begin{align}
\begin{split}
\bm{0}&=\left(c_0  \nabla^\top_{\bm{\dot{q}}}\nabla_{\bm{\dot{q}}}\phi_{0}\right) \bm{\ddot{q}} + \left(c_0  \nabla^\top_{\bm{q}}\nabla_{\bm{\dot{q}}}\phi_{k}\right) \bm{\dot{q}} \\ &-\left(c_0 \nabla_{\bm{q}}\phi_{k}\right) \\ 
&= c_0\left( \nabla^\top_{\bm{\dot{q}}}\nabla_{\bm{\dot{q}}}\phi_{k}\bm{\ddot{q}} + \nabla^\top_{\bm{q}}\nabla_{\bm{\dot{q}}}\phi_{k}\bm{\dot{q}} - \nabla_{\bm{q}}\phi_{k} \right) .
\end{split}
\end{align}

As it can be seen from the equation above, even though we have three different terms in the equation of motion, all of them correspond to the same coefficient $c_0$. This is not the case with SINDy-PI as it would require three different coefficients to represent different terms in the equations of motion. Hence, with fewer parameters, it would be easier to learn the model with xL-SINDy than SINDy-PI while maintaining the same amount of coverage of possible terms in the actual equations of motion. This is one of the reasons why xL-SINDy is more robust than SINDy-PI.

Another reason why xL-SINDy is more robust than SINDy-PI is how the learning process is done. SINDy-PI uses the sequential threshold least-square method \cite{Kaheman2020} where the basic idea is to run the least square method and remove terms with low coefficient sequentially (hard-thresholding). However, we experimentally found that the sequential least-square method often fails to remove non-relevant terms if the training data is corrupted. Instead, we combine the idea of hard-thresholding from the sequential least-square method and soft-thresholding from Lasso regression with the proximal gradient method. We found that it performs better to remove non-relevant terms with a higher level of noise. As it can be seen in the table \ref{table-num-library}, in the case of the double pendulum and spherical pendulum, even though xL-SINDy has more terms in the library than SINDy-PI, xL-SINDy still performs better than SINDy-PI with a higher level of noise. 

Finally, SINDy-PI may have a problem when the incorrect combination of denominator terms is discovered. In rational function, when the denominator is equal to zero, its value blows up. Indeed, this is the case of SINDy-PI in the case of the cart pendulum and the spherical pendulum from our experimental results. On the other hand, xL-SINDy only contains rational terms since it is a Lagrangian mechanic function. So, it would minimize the possibility of a model that blows up due to incorrect terms. 

Like other learning-based methods, xL-SINDy introduces several hyperparameters during the learning process, such as the sparsity constrain $\lambda$, the learning rate $\alpha$, the tolerance for the cost function, and the cut-off threshold in the hard-thresholding process. Tuning the hyperparameters is also vital for the learning outcome, especially the initial value of learning rate $\alpha$ and sparsity constraint $\lambda$. The process of hyperparameter tuning is currently done manually with trial and error. 

One major limitation of xL-SINDy is the difficulty in designing the library. Prior knowledge of the systems is essential to deciding what candidate functions we should include in the library. A large number of candidate functions in the library are more likely to be sufficient, but it makes the sparse optimization more challenging and less robust against noise \cite{Kaheman2020}. Hence, balancing this trade-off is crucial for the outcome of the learning process. 

A better mechanism to handle a large library is crucial in applying xL-SINDy to more complex systems with a higher degree of freedom. One possible way to tackle a large number of the library is by the library-bootstrapping method as in the case of Ensemble-SINDy \cite{Fasel2022}. Many smaller libraries are created by sampling the terms without replacement from the original library, and several different models are learned separately. Once the learning process is done, all terms with low probability inclusion are then removed. This process can be repeated until sufficient results are obtained.  

So far, we have not considered the presence of external non-conservative force acting on the systems. In real-world scenarios, no matter how small, non-conservative forces such as damping or friction are always present. Taking into account this external force in the model is of importance to make xL-SINDy applicable to real systems. One possible way to incorporate non-conservative force is by using the generalized Rayleigh's dissipation function \cite{Minguzzi2015}. Like the Lagrangian, Rayleigh's dissipation function is a single scalar quantity and can be incorporated into Euler-Lagrange's equation. We can model the generalized Rayleigh's dissipation function as a linear combination of candidate functions and learn both Lagrangian and Rayleigh's dissipation function simultaneously.

\section{Conclusion}
In this work, we developed a method called extended Lagrangian-SINDy (xL-
SINDy) that can discover the true form of Lagrangian of nonlinear dynamical
systems from noisy measurement data. We model the Lagrangian as a linear combination of nonlinear candidate functions and use Euler-Lagrange's equation to formulate the objective cost function. We use the proximal gradient method to optimize the cost function and obtain sparse expression of Lagrangian. We demonstrated the effectiveness of xL-SINDy and showed that xL-SINDy is more robust against noise compared to other methods. 

It is worth noting that our proposed method out-performs SINDy-PI (parallel, implicit), a recent robust variant of SINDy developed for implicit dynamics and rational nonlinearities. The robustness against noise was improved by 8-20 times compared to SINDy-PI. We believe that xL-SINDy is a promising approach for the identification of interpretable models of nonlinear dynamics. The focus of our next work is to consider non-conservative forces in the model and apply xL-SINDy to real systems.

\addtolength{\textheight}{-12cm}   

\bibliographystyle{IEEEtran}
\bibliography{ref}

\end{document}